\documentstyle[aps,prl,twocolumn]{revtex}
\begin{document}
\draft
\title{Theory of nonlinear Landau-Zener tunneling}
\author{Jie Liu$^{1,2}$, Li-Bin Fu$^2$, Bi-Yiao Ou$^2$, Shi-Gang Chen$^2$,
and Qian
Niu$^1$ }
\address{$^1$Department of Physics, University of Texas, Austin, Texas
78712\\
$^2$Institute of Applied Physics and Computational Mathematics\\
P.O.Box.8009, 100088 Beijing, China}
\maketitle
\date{}

\begin{abstract}
A nonlinear Landau-Zener model was proposed recently to describe, among a
number of applications, the nonadiabatic transition of a
Bose-Einstein condensate between Bloch bands. Numerical analysis revealed a
striking phenomenon that tunneling occurs even in the adiabatic limit as the
nonlinear parameter $C$ is above a critical value equal to the gap $V$ of
avoided crossing of the two levels. In this paper, we present analytical
results that give quantitative account of the breakdown of adiabaticity by
mapping this quantum nonlinear model into a classical Josephson Hamiltonian.
In the critical region, we find a power-law scaling of the nonadiabatic
transition probability as a function of $C/V-1$ and $\alpha $, the crossing
rate of the energy levels. In the subcritical regime, the transition
probability still follows an exponential law but with the exponent changed
by the nonlinear effect. For $C/V>>1$, we find a near unit probability for
the transition between the adiabatic levels for all values of the crossing
rate.
\end{abstract}




\section{Introduction}

It is a common practice in the study of quantum systems to consider only a
finite number of energy levels which are strongly coupled. The special case
of truncation to two relevant levels is of enormous practical interest, and
a vast amount of literature has been devoted to the dynamical properties of
the two-level systems \cite{GH98}. The Landau-Zener tunneling between energy
levels is a basic physical process \cite{wn1}, and has wide applications in
various systems, such as current driven Josephson junctions \cite{wn2},
atoms in accelerating optical lattices \cite{wn3}, and field-driven
superlattices \cite{wn4}.

A nonlinear two-level system may arise in a mean field treatment of a
many-body system where the particles predominantly occupy two energy
levels, where the level energies depend on the occupation of the levels due
to the interactions between the particles. Such a model arises in the study
of the motion of a small polaron \cite{polar}, for a Bose-Einstein
condensate in a double-well potential \cite{dwpp1,dwpp2,dwpp3} or in an
optical lattice \cite{becol1,becol2,becol3}, or for a small capacitance
Joseph-junction where the charging energy may be important. In contrast to
the linear case, the dynamical property of a nonlinear two-level model is
far from fully understood, and many novel features are revealed recently 
\cite{wuniu,add}, including the discovery of a nonzero Landau-Zener
tunneling probability even in the adiabatic limit when the nonlinear
parameter exceeds a critical value.

In this paper, we present a comprehensive theoretical analysis of the
nonlinear Landau-Zener tunneling, obtaining analytical results for the
tunneling behavior in various regimes. This is made possible by the
observation that the population difference and the relative phase of the two
levels form a pair of canonically conjugate variables of a classical
Hamiltonian, the Josephson Hamiltonian\cite{dwpp2,add}. The fixed points of this
classical system correspond to the eigenstates of the nonlinear two-level
Hamiltonian. Adiabatic evolution of the fixed points as a function of level
bias correspond to adiabatic evolution of the two eigenstates. The breakdown
of adiabaticity occurs when a fixed point turns into a homoclinic orbit
which is possible if and only if the nonlinear parameter exceeds a critical
value. Then the transition probability between the two energy levels can be
nonzero even in the adiabatic limit, and is found to rise as a power law of
the nonlinear parameter minus its critical value. Right at the critical
value of the nonlinear parameter, the transition probability goes to zero as
a power law of the crossing rate of the two levels. Below the critical
point, the transition probability follows an exponential law as in the
linear case but with the exponent changed due to the nonlinearity. Far above
the critical point, we find a near unit probability of transition between
the adiabatic levels for all values of the crossing rates.

Our paper is organized as follows. In Sec. II we introduce the nonlinear
two-level model and discuss the behavior of its eigenstates and
eigenenergies. In Sec. III, we cast the nonlinear two-level system into the
Josephson Hamiltonian and study the evolution of the fixed points as
functions of the system parameters. In Sec. IV we derive a formula for the
tunneling probability in the adiabatic limit and analyze its power law
behavior above and near the critical value of the nonlinear parameter. In
Sec. V, we calculate the nonadiabatic tunneling probability in the critical
and subcritical regimes. In Sec. VI, we derive the nonadiabatic tunneling
probability for the ultrastrong nonlinear coupling using the stationary
phase approximation. In sec. VII, we draw our conclusions and discuss how
our findings may be observed experimentally.

\section{The nonlinear two-level model}

Our model system consists of two levels as in the standard Zener model but
with an additional energy difference depending on the population in the
levels. It is described by the following Hamiltonian matrix 
\begin{equation}  \label{b}
H(\gamma )=\left( \matrix {\ \frac \gamma 2+\frac C2(|b|^2-|a|^2) & \frac V2 %
\cr \frac V2 & -\frac \gamma 2-\frac C2(|b|^2-|a|^2) } \right) ,
\end{equation}
where $a$ and $b$ are the probability amplitudes. The Hamiltonian is
characterized by
three constants: the coupling between the two levels $V$, the level bias $%
\gamma$ as in the linear Zener model, and the nonlinear parameter $C$
describing the level energy dependence on the populations. The amplitudes
satisfy the Schr\"odinger equation 
\begin{equation}  \label{a}
i\frac \partial {\partial t}\left( \matrix {\ a \cr b } \right) =H(\gamma
)\left( \matrix {\ a \cr b } \right),
\end{equation}
according to which, the total probability $|a|^2+|b|^2$ is conserved and is
set to be $1$.

We wish to study how the system evolves when the level bias $\gamma$ changes
slowly from $-\infty$ to $+\infty$. It will be useful to find the adiabatic
levels by diagonalizing the Hamiltonian (\ref{b}), i.e., solving the
following eigen-equations with an eigenenergy $\epsilon ,$ 
\begin{equation}
(\frac \gamma 2+\frac C2(|b|^2-|a|^2))a+\frac V2b=\epsilon a,  \label{ein1}
\end{equation}
\begin{equation}
\frac V2a-(\frac \gamma 2+\frac C2(|b|^2-|a|^2))b=\epsilon b,  \label{ein2}
\end{equation}
For a solution with nonzero amplitudes, we impose the determinental
condition 
\begin{equation}
\det \left( \matrix {\ \frac \gamma 2+\frac C2(|b|^2-|a|^2)-\epsilon & \frac 
V2 \cr \frac V2 & -\frac \gamma 2-\frac C2(|b|^2-|a|^2)-\epsilon } \right)
=0,  \label{ein3}
\end{equation}
which leads to a quartic algebraic equation for the eigenenergy, 
\begin{equation}
\epsilon ^4+C\epsilon ^3+(\frac{C^2}4-\frac{V^2}4-\frac{\gamma ^2}4)\epsilon
^2-\frac{V^2}4C\epsilon -\frac{V^2C^2}{16}=0.  \label{eq}
\end{equation}
The eigenenergies $\epsilon $ correspond to the real roots of the above
quartic equation. It is readily found that this quartic equation has two
real roots when $C< V$, while there are four real roots when $C>V$. Two
typical parameters are chosen to demonstrate the two cases in Fig.1. At $%
C/V=0.5$(Fig.1a), there are two energy levels corresponding to the two real
roots; At $C/V=2$(Fig.1b), a loop appears at the tip of the lower level in
the regime $-\gamma _c\leq \gamma \leq \gamma _c$, where Eq. (\ref{eq}) has
four real roots (this can be easily verified from the above equation at $%
\gamma =0$). The relation between the singular point $\gamma _c$ and the
parameters will be given by Eq. (\ref{rc}) in the following section.

The eigenstates can be determined at the same time from the above equations, 
\begin{equation}
a=\sqrt{\frac 12+\frac \gamma {2C+4\epsilon }},  \label{estate2}
\end{equation}
\begin{equation}
b=\frac{V\sqrt{2C+4\epsilon }}{4\epsilon \sqrt{C+2\epsilon +\gamma }}.
\label{estate3}
\end{equation}
These eigenstates are not orthogonal to each other for finite $\gamma $, but
become so in the limits of $\gamma \to \pm \infty $, where $\epsilon \to \pm
|\gamma |/2$. For instance, for the lower level, we have $(a,b)\to (1,0)$ at 
$\gamma \to -\infty $ and $(a,b)\to (0,1)$ at $\gamma \to +\infty $. Due to
the nonlinearity of the model, there are two more eigenstates for $C>V$ and
in the region $-\gamma _c<\gamma <\gamma _c$.  These correspond to the loop 
structure in the adiabatic energy levels shown in Fig.1b.   

\section{The classical analog}

In this section, we cast the nonlinear two-level system into an effective
classical system, i.e. the Josephson Hamiltonian. The dynamical mechanism
for the breakdown of the adiabatic evolution will be revealed by
investigating this classical Hamiltonian system. The Schr\"odinger equation (%
\ref{a}) has the following explicit form 
\begin{equation}
i\frac \partial {\partial t}a=(\frac \gamma 2+\frac C2(|b|^2-|a|^2))a+\frac V%
2b,  \label{a1}
\end{equation}
\begin{equation}
i\frac \partial {\partial t}b=-(\frac \gamma 2+\frac C2(|b|^2-|a|^2))b+\frac 
V2a,  \label{a2}
\end{equation}
which describes the evolution of the amplitudes for general occupations and
phase of the wavefunctions. The physically interesting information of the
state is completely represented by the population difference 
\begin{equation}
s(t)=|b|^2-|a|^2,  \label{bal}
\end{equation}
and the relative phase 
\begin{equation}
\theta (t)=\theta _b(t)-\theta _a(t).  \label{pha}
\end{equation}
From Eqs. (\ref{a1}) and (\ref{a2}), we can obtain the following equations
of motion 
\begin{equation}
\frac d{dt}s(t)=-V\sqrt{1-s^2(t)}\sin\theta (t),  \label{ceq1}
\end{equation}
\begin{equation}
\frac d{dt}\theta (t)=\gamma +Cs(t)+\frac{Vs(t)}{\sqrt{1-s^2(t)}}\cos \theta
(t).  \label{ceq2}
\end{equation}
These equations can be rewritten in the canonical form $\dot s=-\frac{%
\partial H_e}{\partial \theta }, \dot \theta =\frac{\partial H_e}{\partial s}
$, with an effective classical Hamiltonian given by 
\begin{equation}
H_e(s,\theta ,\gamma )=\frac C2s^2(t)+\gamma s(t)-V\sqrt{1-s^2(t)}\cos
\theta (t),  \label{hamil}
\end{equation}
which has the form of a Josephson Hamiltonian.

Since we are interested in the situation of a quasistatic level bias $\gamma$%
, it will be useful to find the fixed points of the classical dynamics when $%
\gamma$ is fixed in time. The fixed points correspond to the eigenstates of
the nonlinear two-level system, and are obtained by equating the right hand
sides of eqs.(\ref{ceq1}) and (\ref{ceq2}) to zero, yielding 
\begin{equation}
\theta ^{*}=0,\;\pi ,  \label{tmp1}
\end{equation}
\begin{equation}
\gamma +Cs^{*}+\frac{Vs^{*}}{\sqrt{1-{s^{*}}^2}}\cos{\theta ^{*}}=0,
\label{tmp2}
\end{equation}
These can be consolidated into a single quartic equation 
\begin{equation}
{s^{*}}^4+2\frac \gamma C{s^{*}}^3-(1-\frac{\gamma ^2}{C^2}-\frac{V^2}{C^2}){%
s^{*}}^2-2\frac \gamma Cs^{*}-\frac{\gamma ^2}{C^2}=0.  \label{fixed}
\end{equation}

The number of the fixed points depends on the nonlinear parameters $C$. For
weak nonlinearity, $C/V<1$, only two fixed points exist, denoted as $P_1$
and $P_2$ in the panels of Fig.2. These are elliptic fixed points
corresponding to the maximum and minimum of the classical Hamiltonian, and
each are surrounded by closed (elliptic) orbits. The fixed points are
located on the lines of $\theta ^{*}=\pi $ and $0$, meaning that the two
corresponding eigenstates of the two level system have $\pi $ and $0$
relative phases. As the level bias changes from $\gamma =-\infty $ to $%
+\infty $, $P_1$ moves smoothly along the line $\theta ^{*}=\pi $ from the
bottom $(s=-1)$ to the top $(s=+1)$, corresponding to the lower branch of
the energy levels in Fig.1(a). On the other hand, $P_2$ moves from the top
to the bottom corresponding to the upper branch.

For stronger nonlinearity, $C/V>1$, two more fixed points can appear when
the level bias lies in a window $-\gamma _c<\gamma <\gamma _c$. The boundary
of the window can be obtained from the condition of real roots for quartic
equations, yielding 
\begin{equation}
\gamma _c=(C^{2/3}-V^{2/3})^{3/2}.  \label{rc}
\end{equation}
As shown in Fig.3(c-e), both of the new fixed points lie on the line $\theta
^{*}=\pi $, one being elliptic ($P_4$) as the original two and one being
hyperbolic ($P_3$) corresponding to a saddle point of the classical
Hamiltonian. One of the original fixed point, $P_2$, still moves smoothly
with $\gamma $, corresponding to the upper adiabatic level in Fig.1(b). The
other, $P_1$, moves smoothly up to $\gamma =\gamma _c$, where it collides
with $P_3$, corresponding to the branch $OXT$ of the lower level in
Fig.1(b). The new elliptic point $P_4$, created at $\gamma =-\gamma _c$
together with $P_3$, moves up to the top, corresponding to the branch $WXM$
of the lower level. The hyperbolic point $P_3$, moves down away from its
partner after creation and is annihilated again with $P_1$ at $\gamma
=\gamma _c$, corresponding to the top branch $WT$ of the lower level. The
eventual fate of the fixed point $P_1$ will determine the adiabatic
tunneling probability as shown below.

\section{Adiabatic tunneling due to nonlinearity}

For quasistatic change of the level bias $\gamma $, a closed orbit in the
classical dynamics remains closed such that the action 
\begin{equation}
I=\frac 1{2\pi }\oint sd\theta ,  \label{action}
\end{equation}
is invariant in time according to the classical adiabatic theorem \cite{ldm}%
, which is valid as long as the relative change of the system parameter in a
period of the orbit is small, i.e. 
\begin{equation}
Td\gamma /dt\ll \gamma .  \label{cond}
\end{equation}
The action equals the phase space area enclosed by the closed orbit, and is
zero for a fixed point which has no area. Since the closed orbits
surrounding an elliptic fixed point all have finite periods $T$, they should
evolve adiabatically with the area of each fixed in time. We thus expect an
elliptic fixed point to remain as a fixed point during the quasistatic
change of the system parameter. For the case of $C/V<1$, the two fixed
points (both elliptic) evolve adiabatically throughout the entire course of
change in the level bias, implying the absence of transition between the
eigenstates in the adiabatic limit. This is still true for the fixed point $%
P_2$ in the case of $C/V>1$, meaning a state starting from the upper level
will remain in the upper level.

The adiabatic condition is broken, however, when $P_1$ collides with the
hyperbolic fixed point $P_3$ to form a homoclinic orbit where the period $T$
diverges. Nevertheless, the classical 'particle' will remain on this orbit,
because the orbit is surrounded from both outside and inside by closed
orbits of finite periods, which should form adiabatic barriers to prevent
the particle from escaping. After this crisis, the homoclinic orbit turns
into an ordinary closed orbit of finite period, and will evolve
adiabatically for $\gamma>\gamma_c$ according to the rule of constant action
which is now nonzero. In this region of the level bias, the population
difference oscillates along the orbit, which is just the feature described
in Ref.\cite{wuniu} that ``there is a violent shaking above and about the
lower branch of the adiabatic level after the terminal point''.

The adiabatic tunneling probability can then be calculated from the action $%
I_c$ of the homoclinic orbit. This orbit eventually evolves into a straight
line at $s=s_f$ (see Fig.4), where the condition of constant action demands
that 
\begin{equation}
s_f=1-I(s_c).  \label{sss}
\end{equation}
On the other hand, from the definition of population difference, we find the
final probabilities to be given by 
\begin{equation}
\left( \matrix {\ |a_f|^2 \cr |b_f|^2 } \right) =\left( \matrix {\ \frac{%
1-s_f}2 \cr \frac{1+s_f}2 } \right).  \label{est}
\end{equation}
The adiabatic tunneling probability is then expressed as 
\begin{equation}
\Gamma _{ad}=|a_f|^2=I_c/2.  \label{ratee}
\end{equation}

The task we face then is to calculate the action of the homoclinic orbit.
Let $s_c$ be the population difference at the degenerate point $P_c$ where $%
P_1$ collides with $P_3$ when $\gamma =\gamma _c$. It can be obtained by
substituting the expression (\ref{rc}) into Eq. (\ref{fixed}) as 
\begin{equation}
s_c=-\frac{\sqrt{1-(V/C)^{2/3}}}2\left( 1+(V/C)^{2/3}\right) -\frac{\gamma _c%
}{2C}.  \label{pc}
\end{equation}
Considering the fact that the degenerate point $P_c$ lies on $\theta =\pi ,$
we find the total energy (the value of the classical Hamiltonian) as 
\begin{equation}
E_c=\frac C2(s_c)^2+\gamma _cs_c+V\sqrt{1-(s_c)^2}.  \label{ene}
\end{equation}
The homoclinic orbit should have this energy, and its trajectory 
\begin{equation}
s=s(\theta ;E_c),  \label{pf}
\end{equation}
may be obtained by equating the classical Hamiltonian (\ref{hamil}) to $E_c$%
. The corresponding action can then be evaluated to give the tunneling
probability as 
\begin{equation}
\Gamma _{ad}={\frac 12}I(s_c)=\frac 1{4\pi }\oint s(\theta ;E_c)d\theta .
\label{rateee}
\end{equation}
This has been evaluated numerically, and the results compare extremely well
with those obtained by directly integrating the time-dependent nonlinear
Schr\"odinger equation using the Runge-Kutta algorithm (See Fig.5).

The adiabatic tunneling probability can be evaluated analytically in the
critical region of $\delta =C/V-1\to 0$. The singular point of the level
bias is found to leading order as 
\begin{equation}
\gamma _c\simeq V({\frac 23}\delta )^{3/2}.
\end{equation}
The homoclinic orbit is confined near the critical point, with its top at 
\begin{equation}
s_t\simeq s_c+3\sqrt{\frac 23\delta }.
\end{equation}
We may expand the classical Hamiltonian to leading orders of $s-s_c$ and $%
\theta -\pi $ to find 
\begin{equation}
\theta -\pi \simeq \sqrt{2\gamma _c(s-s_c)}+\frac 12\sqrt{2\gamma _c}%
(s-s_c)^{3/2}.
\end{equation}
From the area of this orbit the adiabatic tunneling probability for this
limiting case is found to be given by the power law 
\begin{equation}
\Gamma _{ad}=\int_0^{s_t}(\theta -\pi )ds=\frac 4{3\pi }\delta ^{\frac 32}.
\label{small}
\end{equation}

We may also obtain the adiabatic tunneling probability analytically in
another limit, $C/V\rightarrow \infty$. To leading order in the small
quantity $\sigma=V/C$, we find 
\begin{equation}
\gamma _c \simeq C(1-\frac 32\sigma ^{2/3}),\;
\end{equation}
\begin{equation}
s_{c} \simeq -1+\frac 12\sigma ^{2/3}.
\end{equation}
In this case, the homoclinic orbit is confined near $s=-1$, and the
deviation is the same order as $s_c-(-1)$. If we write $s-(-1)=\eta^2%
\sigma^{2/3}$, then 
\begin{equation}
\frac 34\/-3\ \eta ^2+\eta ^4-2\ \sqrt{2}\ \eta \cos(\theta )\simeq 0.
\end{equation}
By solving this equation and considering $s=\eta ^2(\theta )\sigma ^{2/3}-1,$
the adiabatic tunneling probability can be obtained 
\begin{equation}
\Gamma _{ad}\simeq 1-\frac 32(\frac VC)^{2/3}.  \label{bigt}
\end{equation}

\section{Nonadiabatic tunneling at and below the critical point}

We have thus found that it is possible to tunnel between the adiabatic
levels even in the adiabatic limit if the nonlinear parameter exceeds a
critical value, i.e., $C/V>1$. We now consider the case of nonzero sweeping
rates, and study how the nonliearity affects the nonadiabatic tunneling
probability. In the linear case $C=0$, the Landau-Zener formula prescribes
an exponential tunneling probability between the adiabatic levels, 
\begin{equation}
\Gamma \sim \exp (-\frac{\pi V^2}{2\alpha }).
\end{equation}
For the nonadiabatic case, we will focus our attention in this section to
the critical point and the subcritical region, and consider the near
adiabatic case (i.e., $\alpha \ne 0$ and $\alpha <<1$).

For this purpose, we need to investigate the evolution of the fixed point $%
P_1$ as well as the nearby periodic orbits. Then, in addition to the action
variable $I$ mentioned in the above section, we should introduce its
canonical conjugate -- the angle variable $\phi $. They satisfy the
following differential equations\cite{ldm}, 
\begin{equation}
\dot I=-\frac{\partial R}{\partial \phi }\dot \gamma ,  \label{e53}
\end{equation}
\begin{equation}
\dot \phi =\omega (I,\gamma )+\frac{\partial R}{\partial I}\dot \gamma .
\label{e54}
\end{equation}
The function $R(I,\phi )$ is defined as $\frac \partial {\partial \gamma }%
\int s(\theta ,E,\gamma )d\theta $, where $E$ is the energy of the periodic
orbit. Here, $\omega $ is the frequency of orbits in the immediate
neighborhood of the fixed point $P_1$. Its expression is particular simple
at $I=0$, corresponding to the fixed points $P_1$. It can be calculated by
linearizing the equations of motion eqs.(\ref{ceq1},\ref{ceq2}) near the
fixed point as
\begin{equation}
\omega ^{*}=V\sqrt{1/(1-({s}^{*}{)}^2)-\frac CV\sqrt{1-({s^{*})}^2}}.
\label{e61}
\end{equation}

As in the adiabatic case considered in the last section, the transition
probability is still given by the increment of the action, i.e., 
\begin{equation}
\Gamma =\frac 12\Delta I.  \label{jll}
\end{equation}
The reason is as follows. The initial state at $\gamma=-\infty$ is the fixed
point $P_1$ with zero action, and the final state is an orbit of finite
action which becomes a horizontal line $s=s_f$ at $\gamma=+\infty$. Then an
integration of Eq.(\ref{e53}) yields 
\begin{equation}
\Delta I=-\int_{-\infty }^{+\infty }\frac{\partial R}{\partial \phi}\frac{
d\gamma }{dt}\frac{d\phi}{\dot \phi}.  \label{e56}
\end{equation}

To evaluate this integral, we need to express $\dot \phi$ as a function of $%
\phi$ itself. Near the adiabatic limit, we may omit the second term in Eq.(%
\ref{e54}) and set $\omega(I,\gamma)=\omega^*(\gamma)$ in the first term,
yielding 
\begin{equation}
\dot \phi=V\sqrt{1/(1-({s}^{*}{)}^2)-\frac CV\sqrt{1-({s^{*})}^2}}.
\end{equation}
On the other hand, by substituting $\theta ^{*}=\pi $ into eq.(\ref{tmp2})
and differentiating with respect to time, one gets 
\begin{equation}
\frac{dt}{ds^{*}}=\frac V\alpha \left( 1/\sqrt{1-({s}^{{*}}{)}^2}+({s}^{{*}}{%
)}^2/(1-({s}^{{*}}{)}^2)^{3/2}-C/V\right) .  \label{e63}
\end{equation}
Combining these equations, one may relate $s^*$ to $\phi$ and thus express $%
\dot\phi$ as a function of $\phi$ itself.

The principal contribution to the integral comes from the neighborhood of
the singularities of the integrand, which comes only from the zeros of the
frequency $\dot \phi=\omega^*(\gamma)$. These zero points are easily found
from Eq.(\ref{e61}) as 
\begin{equation}
s_0^{*}=[{1-(V/C)^{\frac 23}}]^{\frac{1}{2}}.  \label{e62}
\end{equation}
As will be shown below, the integral is exponentially small if there is no
real singularities, and becomes a power law in the sweeping rate if there is
a singularity on the real axis.

We first consider the case of critical nonlinearly, $C/V=1$, for which the
singular point occurs at $s^{*}=0$. Near this point, we find from Eqs.(\ref
{e63}) that $\omega^* \simeq \sqrt{\frac 32}Vs^{*}$ and $\phi\simeq \frac 14(%
\frac 32)^{3/2}\frac{V^2}\alpha (s^{*})^4$. Then, we have an approximate
relation $\omega^* \sim \alpha ^{\frac 14}\phi^{\frac 14}$ near the
singularity. Substituting these expressions back to the eq.(\ref{e56}), and
utilizing the fact that ${\frac{\partial R }{\partial \phi}}$ is independent
of $\alpha$, we find a power-law behavior for the tunneling probability 
\begin{equation}
\Gamma \sim \alpha ^{\frac 34}.
\end{equation}
This scaling law has been verified by our numerical calculations (Fig.6).

Now we extend our discussion to nonadiabatic tunneling for subcritical
nonlinearity, where the zeros of the frequency $\omega^*$ are complex.
Because $R$ is a periodic function of the angle variable, it can be expanded
as a Fourier series, 
\begin{equation}
\frac{\partial R}{\partial \phi}=\sum_{l=-\infty }^{+\infty }ile^{il\phi}R_l.
\label{e57}
\end{equation}
After substituting this series into eq.(\ref{e56}), we may deform the
contour of integration into the complex plane. For those positive $l$ terms
in the series, we may raise the contour into the upper half plane until it
is ''caught up'' by a singularity of the integrand, namely a zero of $%
\omega^*$. Let $\phi_0$ be the singularity nearest to the real axis, i.e.
the one with the smallest positive imaginary part. The principal
contribution to the integral comes from the neighborhood of this point,
yielding \cite{ldm}, 
\begin{equation}
\Delta I\sim \exp(-l_s{\rm Im}(\phi_0)),  \label{e58}
\end{equation}
where '${\rm Im}(\phi_0)$' indicates the imaginary part of the complex
number $\phi_0$ and $l_s$ is the lowest order of the Fourier series whose
Fourier component does not vanish. The negative $l$ terms give the same
exponential and so affects only the prefactor which we do not consider here.
In the neighborhood of the elliptic fixed point $P_1$, the system can be
approximated by a harmonic oscillation with the function $R\sim \sin(2\phi)$%
, so we have $l_s=2$.

To find the complex zero $\phi_0$, we combine the equations of (\ref{e63})
to express it as an integral of $s^*$ 
\[
\phi _0=\frac{V^2}\alpha \int_0^{{s_0^{*}}} \sqrt{1/(1-({s}^{{*}}{)}^2)-%
\frac CV \sqrt{1-({s^{*})}^2}} 
\]
\begin{equation}
\left( 1/\sqrt{1-({s}^{{*}}{)}^2}+({s}^{{*}}{)}^2/(1-({s}^{{*}}{)}%
^2)^{3/2}-C/V\right) ds^{*}.  \label{liu2}
\end{equation}
We take the upper bound to be the zero $s^*_0$ of the frequency, while the
lower bound only reflects a certain choice of the origin of $\phi$ which
does not affect the imaginary part.

The tunneling probability is thus found to be proportional to the
exponential 
\begin{equation}
\Gamma \sim \exp (-q\frac{\pi V^2}{2\alpha }),
\end{equation}
where the factor in the exponent is given by 
\begin{equation}
q=\frac 4\pi \int_0^{\sqrt{(V/C)^{2/3}-1}%
}(1+x^2)^{1/4}(1/(1+x^2)^{3/2}-C/V)^{3/2}dx.
\end{equation}
For the linear case $C=0$, the factor $q$ is exactly unit, which means that
the expression of the tunneling probability is identical to the standard
Landau-Zener formula. For the nonlinear case, $C/V>0$, this factor becomes
smaller than one, indicating the enhancement effect on the nonadiabatic
tunneling. As $C/V$ goes up to 1, the critical point, this factor vanishes,
meaning the breakdown of the exponential law.

In Fig.7, we plot this factor and compare it with the results of numerical
integration of the nonlinear Schr\"odinger equation. The agreement is very
well for the nearly linear case, $C/V<<1$, and there is some deviation when $%
C/V$ is close to unity. One reason for the latter is that the prefactor may
be important in such a case. Also, numerical accuracy may be blamed. To
obtain the correct slope in the plot of $ln\Gamma $ vs. $\alpha $, one
should calculate the tunneling probability at infinitesimal $\alpha $.
However, in practical calculations, this parameter is chosen as a finite
value instead of infinitesimal. In \cite{wuniu}, the slope is read in the
regime $\alpha \in [0.02,0.01]$. In our calculations, this range is extended
to $[0.005,0.0025]$. As is well known, in the regime of small $\alpha $, the
numerical fluctuation is serious. This may result in a relative large slope
numerically.

\section{Tunneling at strong nonlinearity}

In an earlier section, we have found that for strong nonlinearity $C/V>>1$,
there is a near unity tunneling probability to the upper adiabatic level
even in the adiabatic limit. This probability can only gets larger when the
sweeping rate is finite. We thus expect that the amplitude $b$ in the
original Schr\"odinger equation remains small all the times, and a
perturbative treatment of the problem becomes adequate. We begin with the
variable transformation 
\begin{equation}
a=a^{\prime }\exp \big(-i\int_0^t(\frac \gamma 2+\frac C2(|b|^2-|a|^2))dt%
\big),
\end{equation}
\begin{equation}
b=b^{\prime }\exp \big(i\int_0^t(\frac \gamma 2+\frac C2(|b|^2-|a|^2))dt\big)%
.
\end{equation}
Then the diagonal terms in Eqs. (\ref{a1},\ref{a2}) are transformed away, and our above
discussions imply that we can take $a^{\prime }\approx 1$, and 
\begin{equation}
b^{\prime }=\frac V{2i}\int_{-\infty }^tdt\exp \big(-i\int_0^t(\gamma
+C(|b|^2-|a|^2))dt\big).
\end{equation}

We need to evaluate the above integral self-consistently. Because of the
large $C$, the nonlinear term in the exponent generally gives a rapid phase
change which makes the integral small. The dominant contribution comes from
the stationary point $t_0$ of the phase where $(-\gamma
+C(1-2|b|^2))_{t_0}=0.$ Expanding about this point, we may write 
\begin{equation}
-\gamma +C(1-2|b|^2)=-\bar \alpha (t-t_0),
\end{equation}
with 
\begin{equation}
\bar \alpha =\alpha +2C[{\frac d{dt}}|b|^2]_{t_0}.  \label{xxx}
\end{equation}
We thus have 
\begin{equation}
|b|^2=(\frac V2)^2\bigg|\int_{-\infty }^tdt\exp (-\frac i2\bar \alpha
(t-t_0)^2)\bigg|^2.
\end{equation}
We can differentiate this expression and evaluate its result at time $t_0$,
obtaining a few standard Fresnel integrals with the result $[{\frac d{dt}}%
|b|^2]_{t_0}=(\frac V2)^2\sqrt{\frac \pi {\bar \alpha }}$. Combining this
with the relation (\ref{xxx}), we come to a closed equation for $\bar \alpha 
$, 
\begin{equation}
\bar \alpha =\alpha +2C(\frac V2)^2\sqrt{\frac \pi {\bar \alpha }}.
\end{equation}

The nonadiabatic transition probability $\Gamma$ is given by 
\[
\Gamma = 1-|b|^2_{+\infty} = 1-(\frac{V}{2})^2\bigg|\int_{-\infty}^{+\infty}
dt \exp(-\frac{i}{2}\bar\alpha(t-t_0)^2)\bigg|^2 
\]
\begin{equation}
=1-\frac{\pi V^2}{2\bar\alpha}.
\end{equation}
Then the above result yields a closed equation for the $\Gamma$, 
\begin{equation}
\frac{1}{1-\Gamma} = \frac{1}{P} + \frac{\sqrt2}{\pi}\frac{C}{V}\sqrt{%
1-\Gamma},
\end{equation}
where $P=\frac{\pi V^2}{2\alpha}$. In the adiabatic limit, i.e. $\frac{1}{P}%
=0,$ we find that $\Gamma = 1-1.7(\frac{V}{C})^{\frac{2}{3}}$. This gives
the same exponent as the exact asymptotic result (\ref{bigt}) obtained in an
earlier section, but the coefficient differs by about 10\%. In the sudden
limit, $\frac{1}{P}\to\infty$, we have $\Gamma = 1- P$, which is exact. In
figure 8, we compare the above analytical results with that from directly
solving the Schr\"odinger equation and show a good agreement.

\section{Conclusions and discussions}

In this paper, we present a theoretical analysis on the Landau-Zener
tunneling in a nonlinear two-level system. Our results can be summarized as:
1) The adiabatic tunneling probability depends only on the ratio between the
nonlinear parameter $C$ and the coupling constant $V$. With increasing the
ratio $C/V$, a transition from zero adiabatic tunneling probability to
nonzero adiabatic tunneling probability occurs at a critical point $C/V=1$.
The critical behavior near the transition point follows a power-law scaling.
2) In the subcritical regime, the nonadiabatic tunneling probability follows
a exponential law, which modifies the Landau-Zener formula by adding a
factor in the exponent. This exponential law breaks down at the critical
point, instead, a power-law shows up. 3)Far above the critical point, we
find a near unit probability of transition between the adiabatic levels for
all values of the crossing rates. The analytical expression is obtained
using the stationary phase approximation.

The experimental observation of these findings is a topic of great interest.
It has been shown that the nonlinear two-level model can be used to
understand the tunneling of a Bose-Einstein condensate between Bloch bands
in an optical lattice \cite{wuniu}. Another physical model for direct
application of our theory is a Bose-Einstein condensate in a double-well
potential\cite{dwpp1,dwpp2,RS99}. The amplitudes of general occupations $%
N_{1,2}(t)$ and phases $\theta _{1,2}$ obey the nonlinear two-mode
Schr\"odinger equations similar to (\ref{a}) (e.g. see \cite{RS99}). After
introducing the new variables $s(t)=(N_2(t)-N_1(t))/N_T$ and $\theta =\theta
_2-\theta _1$, one also obtain a Hamiltonian having the same form as Eq.(\ref
{hamil}) except for the parameters replaced by $V=2K/\hbar ,\gamma
=-[(E_1^0-E_2^0)-(U_1-U_2)N_T/2]/\hbar ,C=(U_1+U_2)N_T/2\hbar . $ Here the
constants have following meanings: Total number of atoms $N_1+N_2=N_T$, is
conserved, $E_{1,2}^0$ are zero-point energy in each well, $U_{1,2}N_{1,2}$
are proportional to the atomic self-interaction energy, and $K$ describes
the amplitude of the tunneling between the condensates.

In practice, such a double well may be achieved by a laser sheet dividing a
trap, and the energy levels may be moved by shifting the laser sheet. Then,
our theoretical results can be directly applied to this system without
intrinsic difficulty. One can design an experiment for observing the
transition to nonzero adiabatic tunneling and verify those scaling law found
in our paper. We hope our discussions will stimulate the experimental works
in the direction.

\section{Acknowledgment}

We thank Prof. X.G. Zhao for useful discussions. This project was supported
by Fundamental Research Project of China, the NSF and the Welch Foundation.

Figure Captions

Fig.1 Adiabatic energy levels for two typical nonlinear cases (solid line).
The dashed lines is the ones for linear case($C=0$). $P_i$ $(i=1,\cdots ,4)$
is the fixed points of $H_e$ system, corresponding to the quantum
eigenstates as shown in (b): $OXT\rightarrow P_1,$ $MXW\rightarrow
P_4,\;WT\rightarrow P_3,$ only $P_3$ is a unstable saddle point, others are
stable elliptic point.

Fig.2 The phase-space portrait of the Hamilton $H_e$ system for various
parameter $\gamma $ at $C/V=0.5$. The arrows refer to the direction of the
shift of the fixed points $P_i$ as $\gamma $ increases. The closed curves
are the periodic trajectories. In this case, no collision between fixed
points occurs, which indicates a zero adiabatic tunneling probability.

Fig.3 The phase-space portrait of the Hamilton $H_e$ system for various
parameter $\gamma $ at $C/V=2$. The arrows refer to the direction of the
shift of the fixed points as $\gamma $ increases. In this case, a collision
occurs at a singular point $\gamma _c$, where $P_2$ and $P_3$ collide and
form a homoclinic orbit with nonzero action. This jump on the action leads
to a nonzero adiabatic tunneling probability.

Fig.4 The homoclinic trajectories associated with the singular points $P_c$
at critical point $\gamma =\gamma _c$ (left column), and its final shape at $%
\gamma =+\infty $ after adiabatic evolution (right column), respectively.
For the action keeps as an invariant during the evolution process, the areas
shadowed are equal for the same $C/V$. Here we show two typical situations
corresponding to smaller and larger nonlinearity, respectively.

Fig. 5 The adiabatic tunneling probability vs. parameter $C/V$. Solid line
is our theoretical results, whereas the symbols-- dot, star and plus denote
the results from solving nonlinear Schr\"odinger equation numerically. Inset
is a magnification of the local plot.

Fig.6 The dependence of the tunneling probability on the parameter $\alpha$
for the critical nonlinearity.

Fig.7 The dependence of the factor $q$ on the ratio $C/V$.

Fig.8 The comparison between our analytical theory and the numerical
simulations in the regime of strong nonlinear coupling.


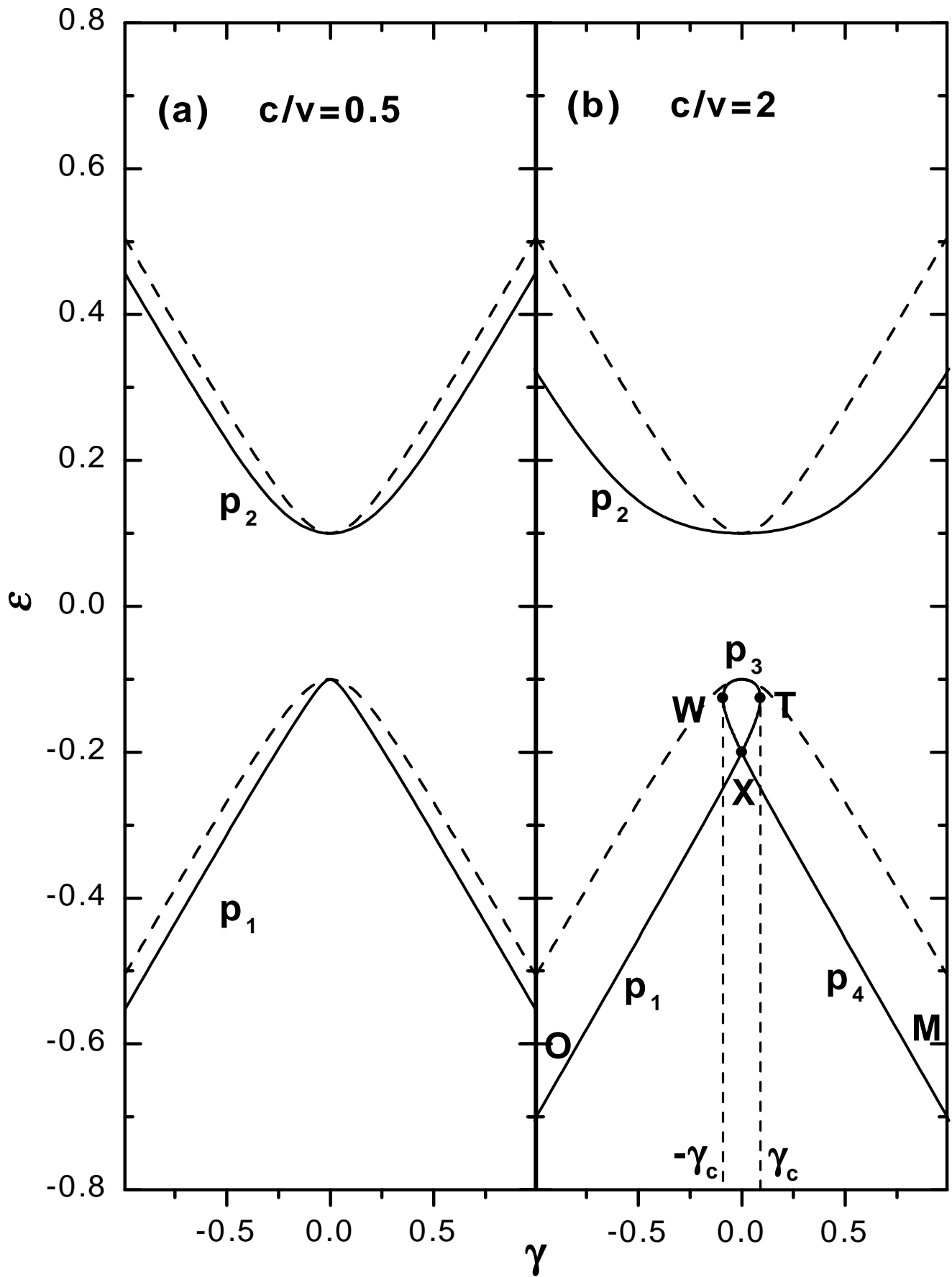
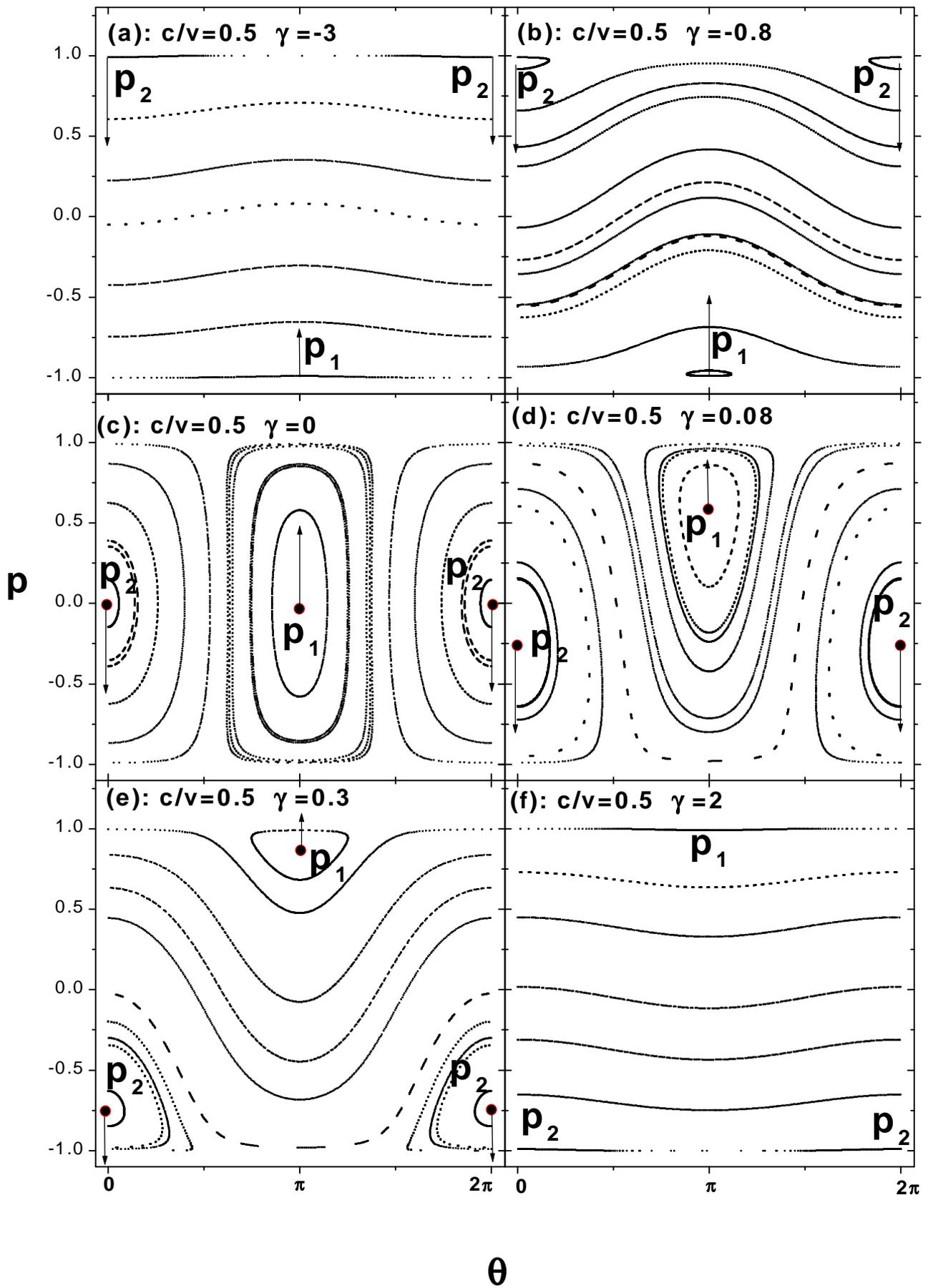
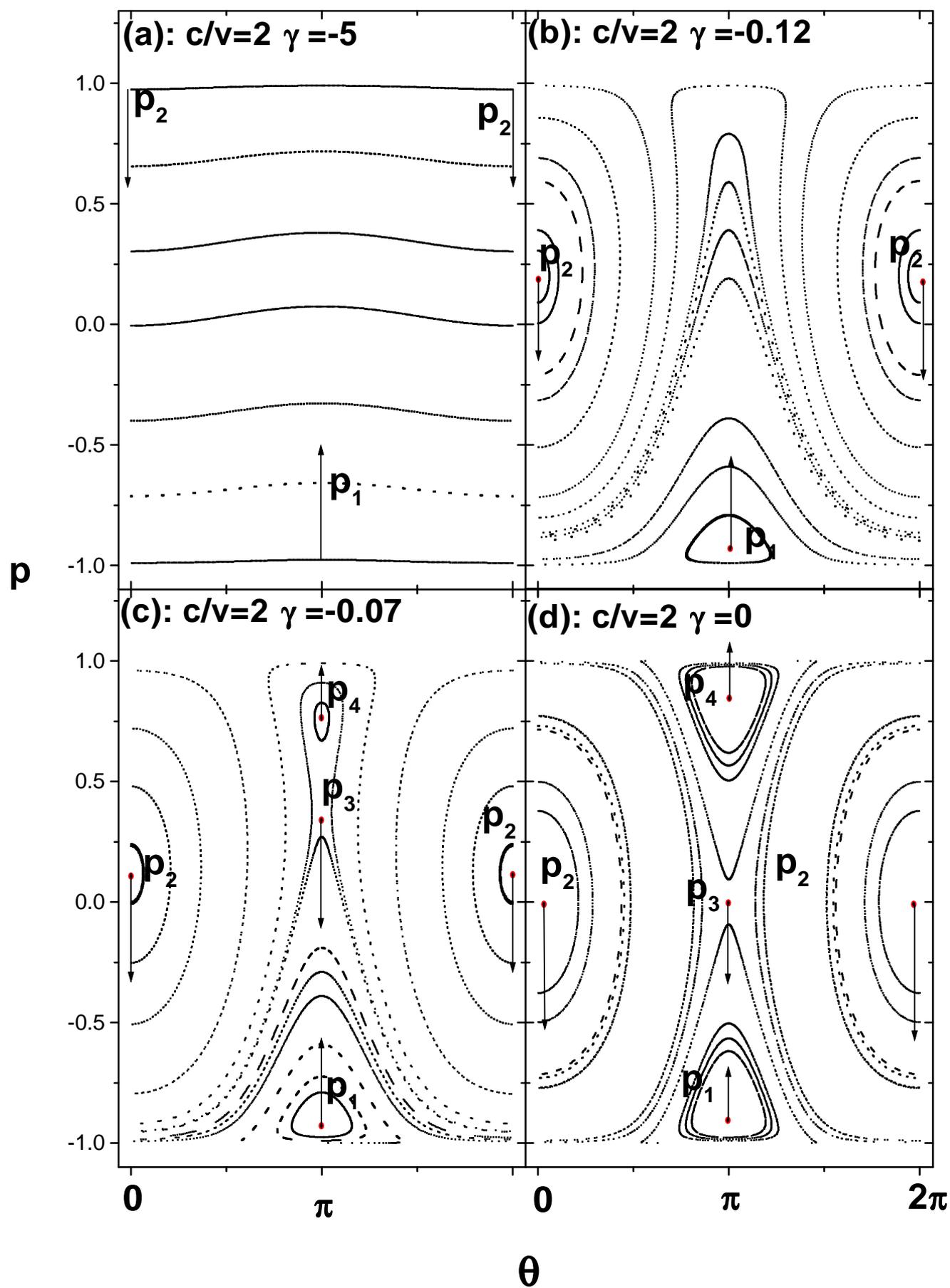
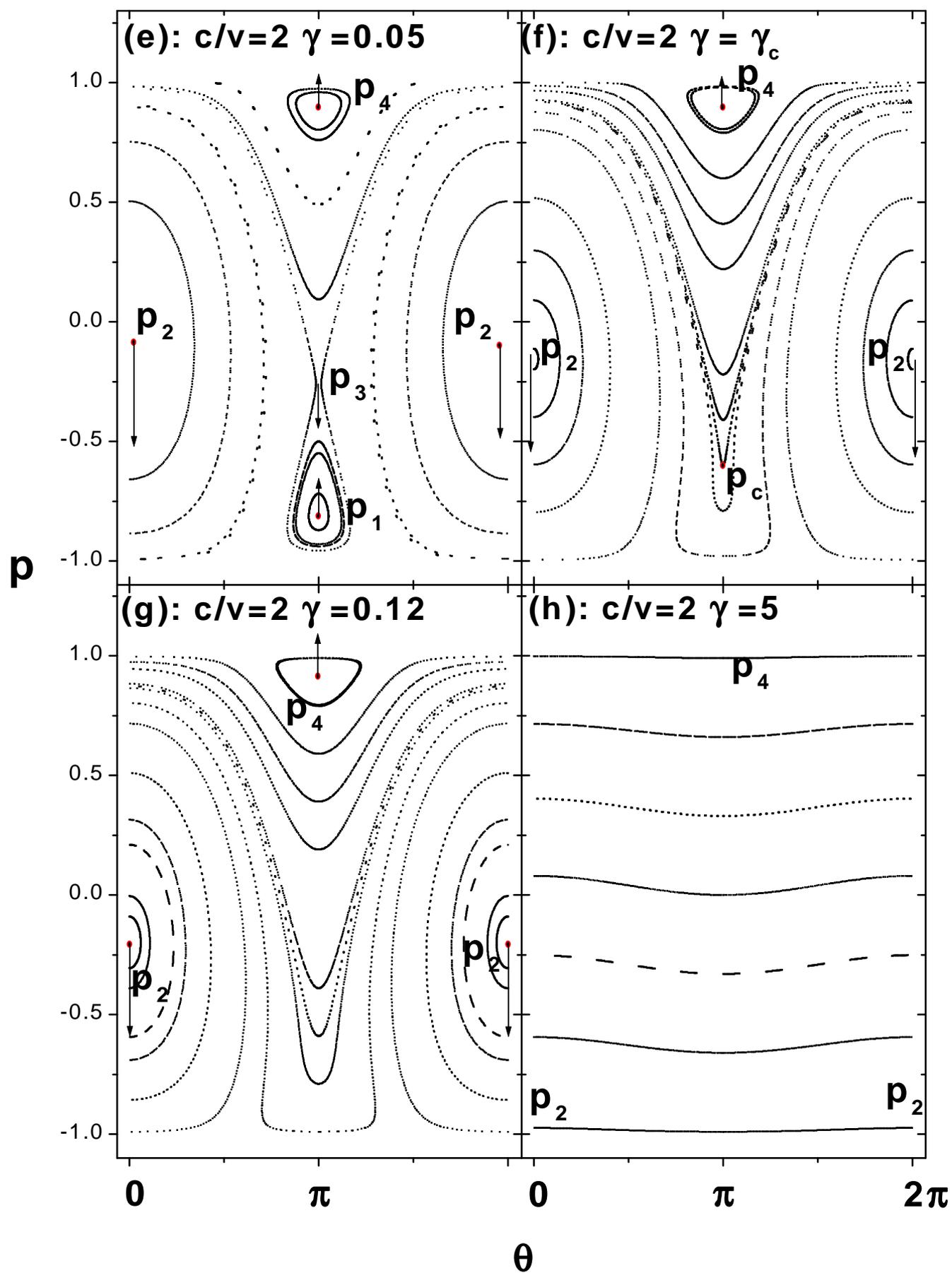
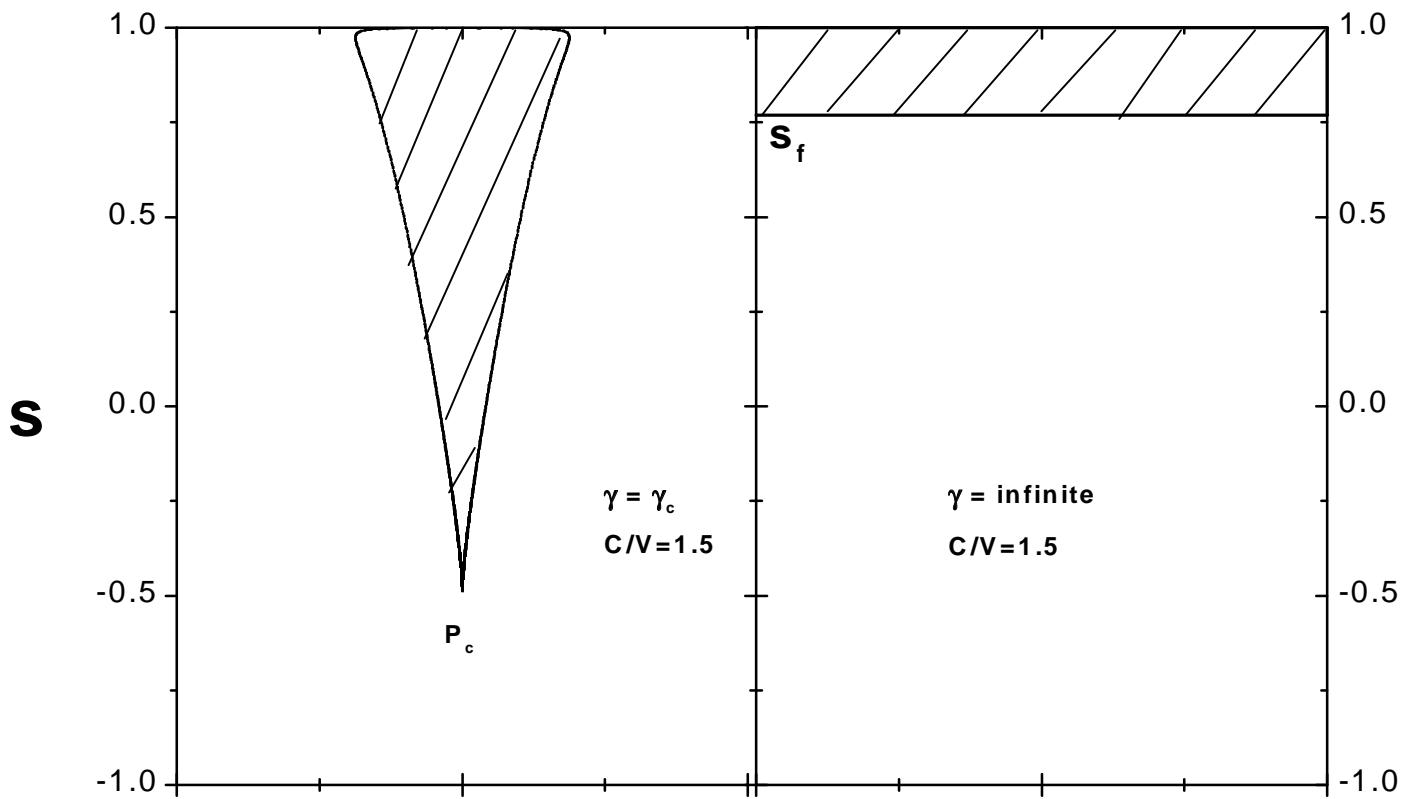
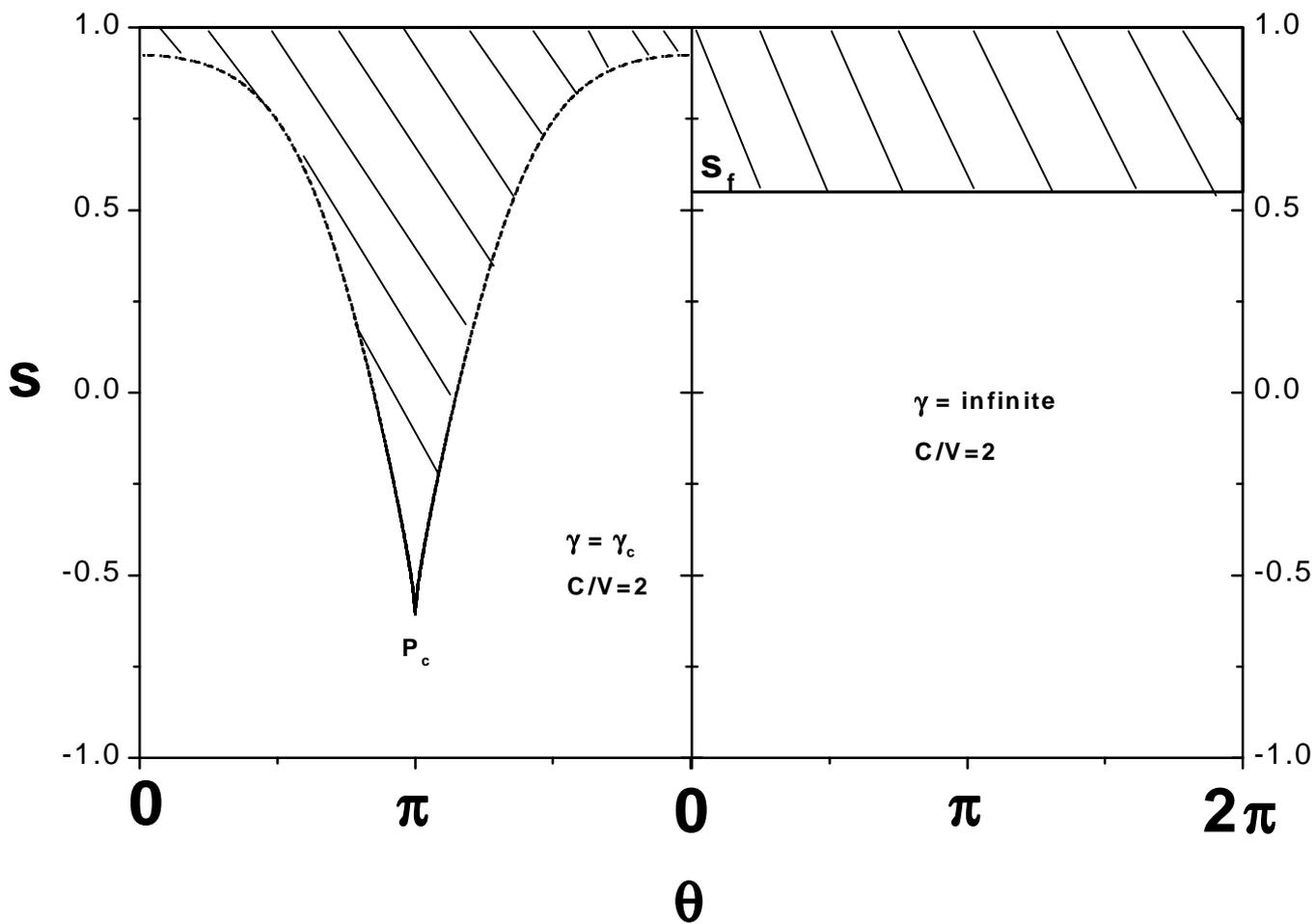
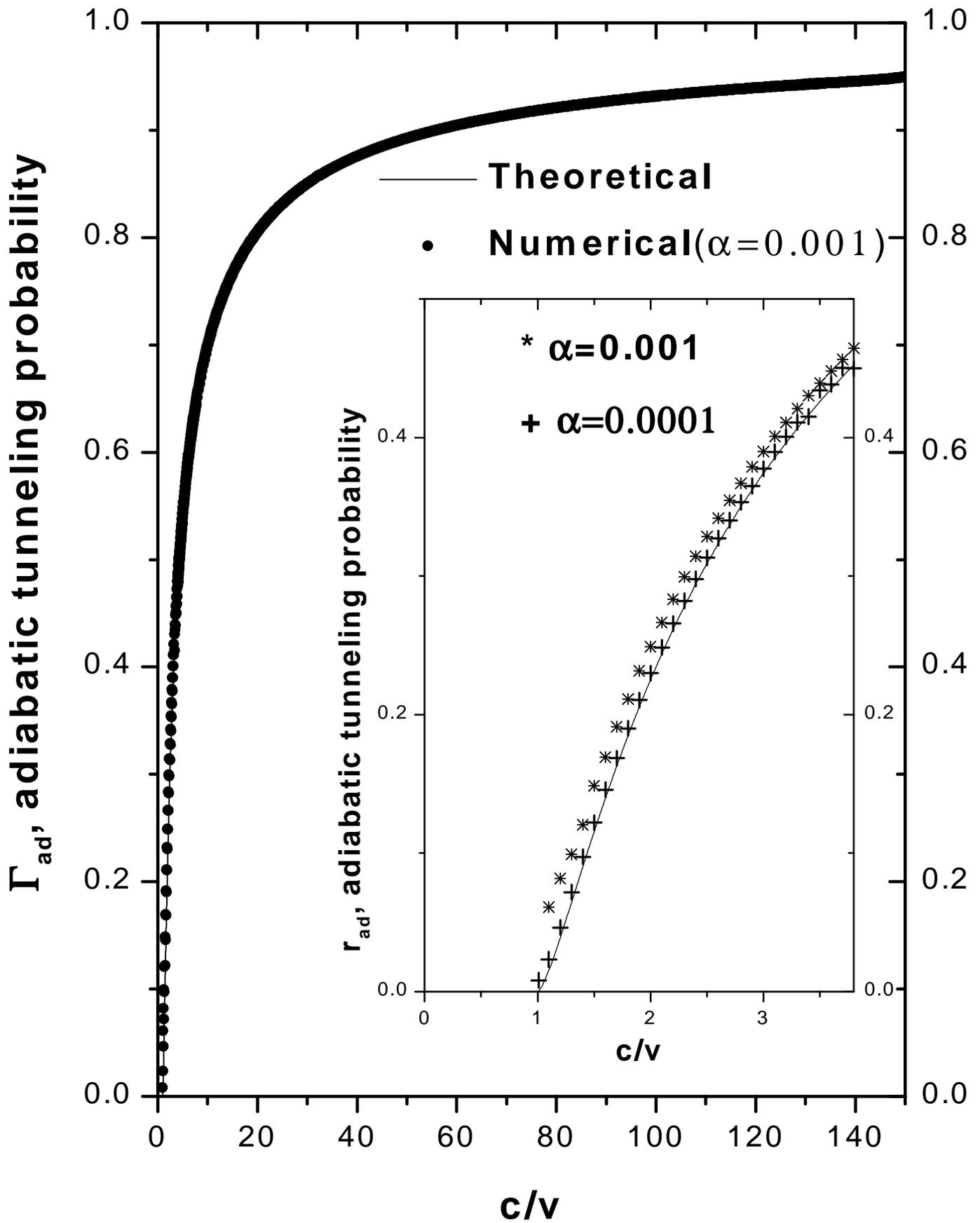
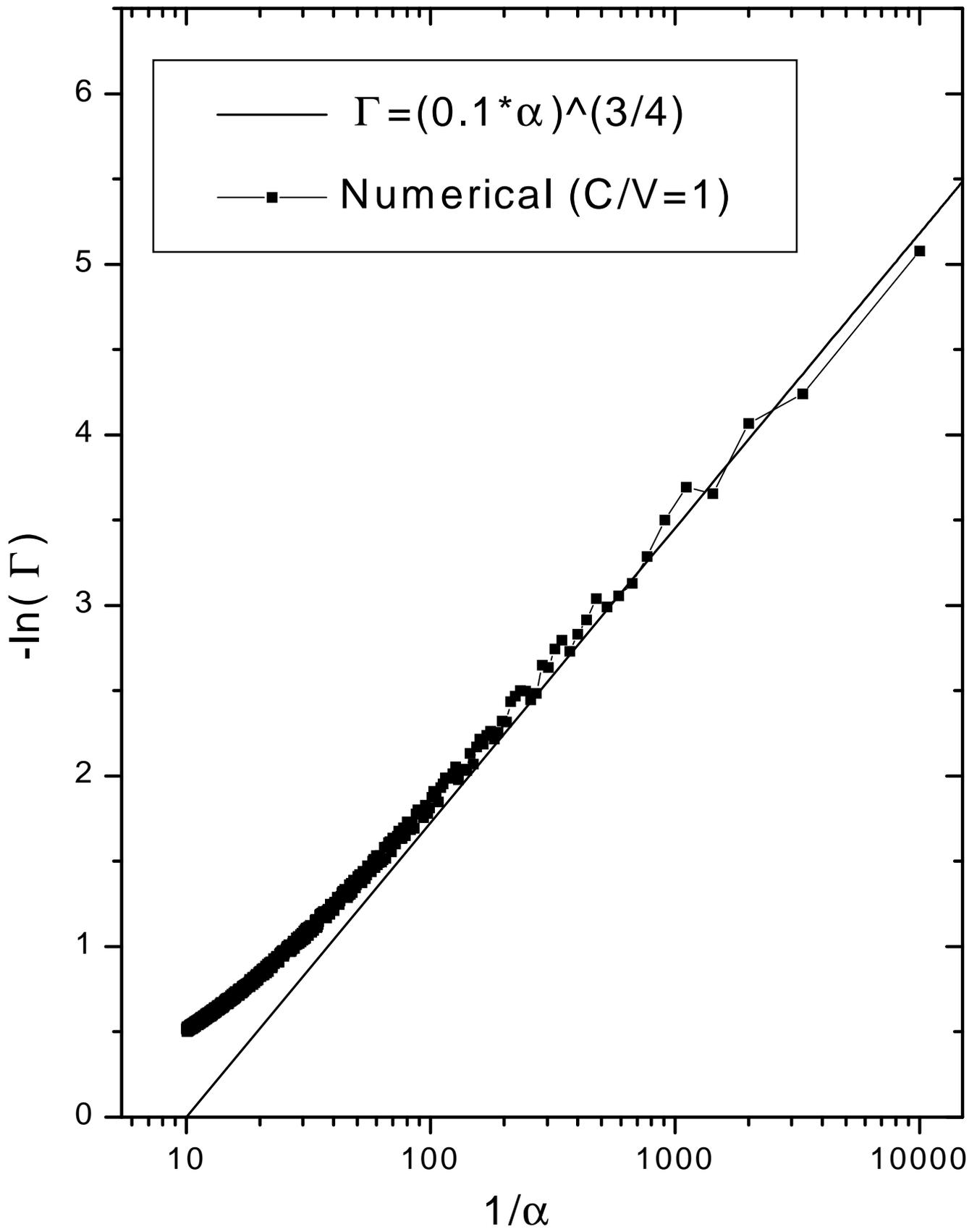
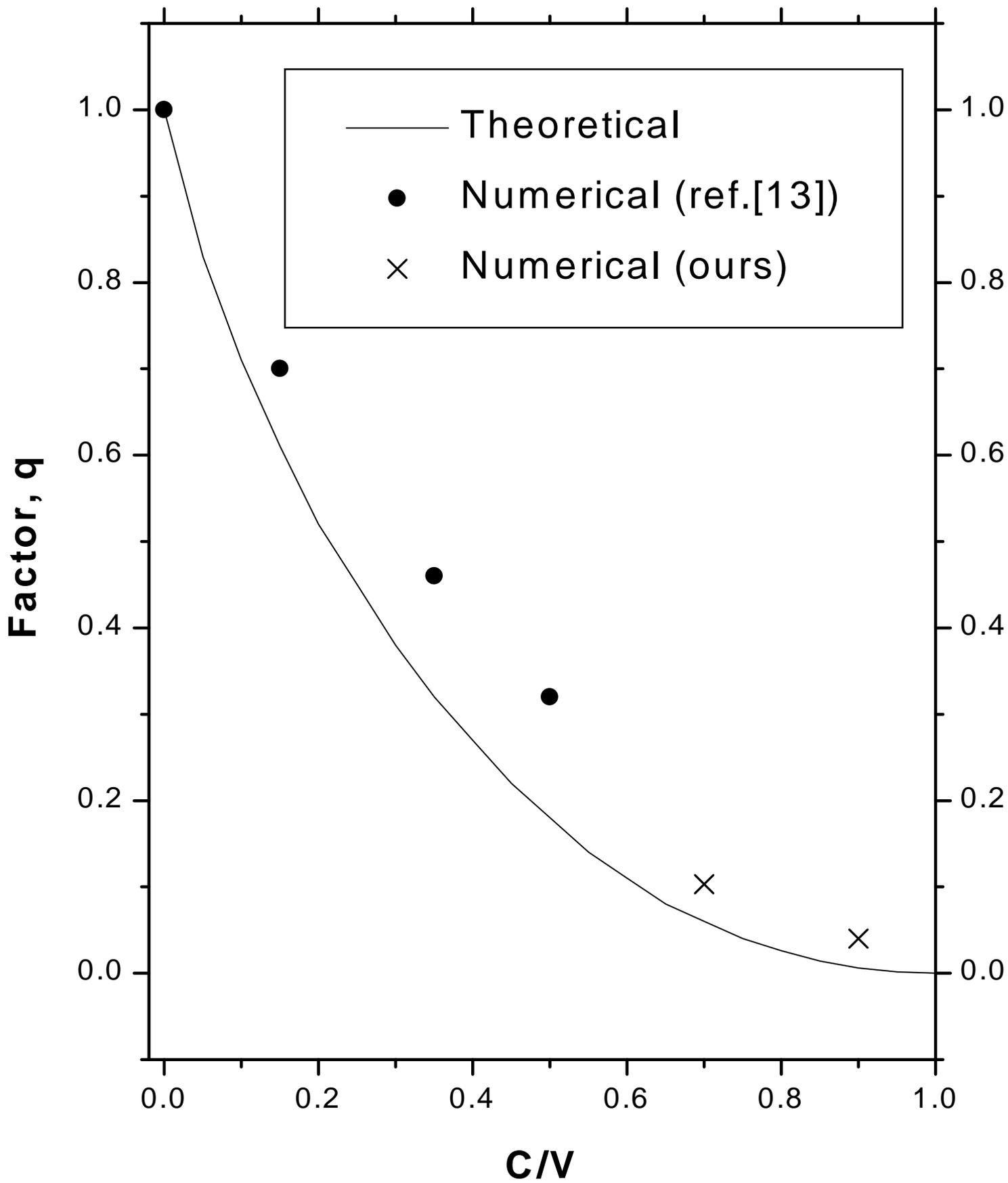
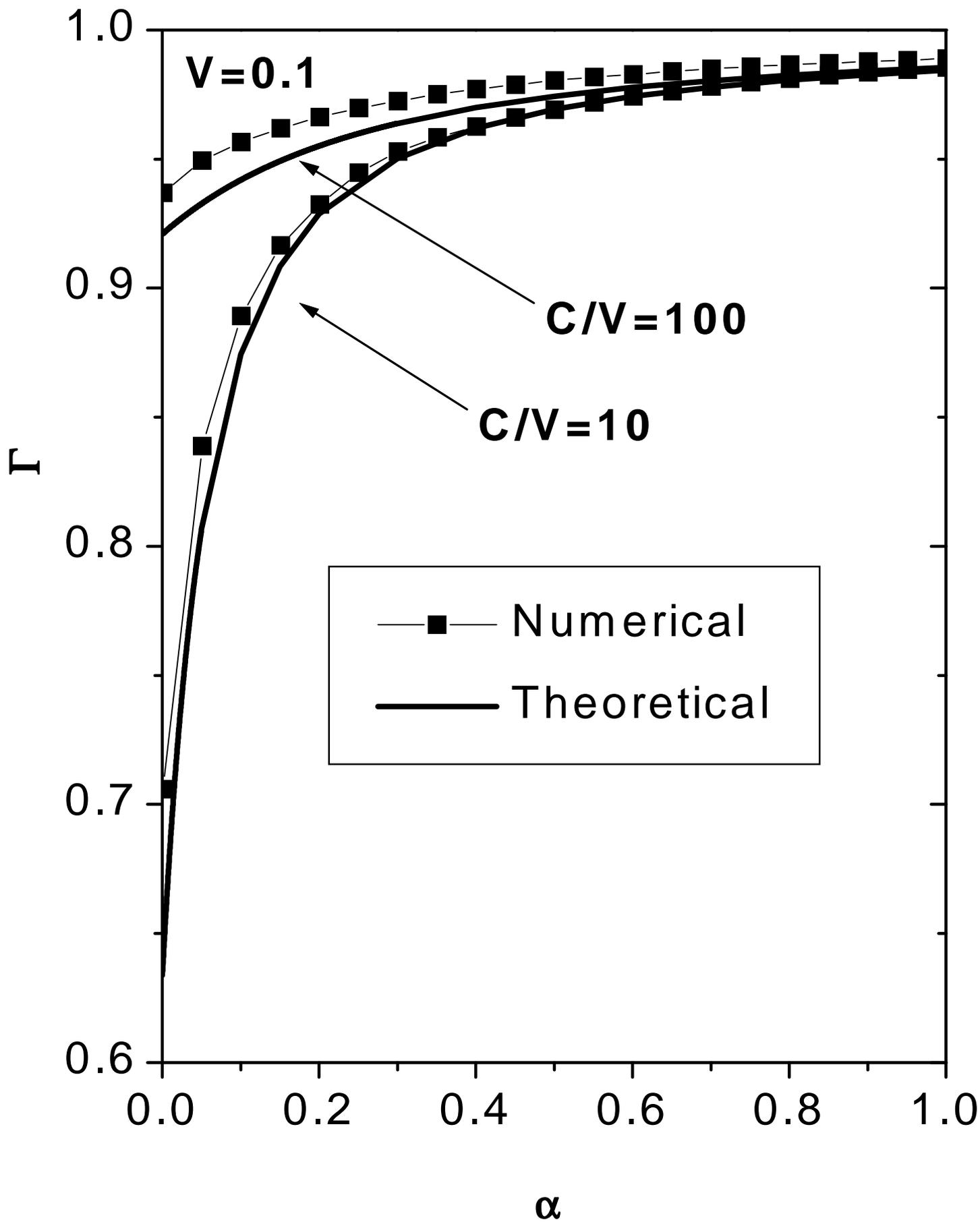
\end{document}